\documentclass[letterpaper,10pt]{article} 
\DeclareRobustCommand{\rchi}{{\mathpalette\irchi\relax}}
\newcommand{\irchi}[2]{\raisebox{\depth}{$#1\chi$}} 

\usepackage{osameet3} 

\usepackage{amsmath,amssymb}
\usepackage[colorlinks=true,bookmarks=false,citecolor=blue,urlcolor=blue]{hyperref} 
\usepackage{wrapfig}
\usepackage{subfig}
\usepackage{floatrow}
\newfloatcommand{capbtabbox}{table}[][\FBwidth]
\renewcommand{\Re}{\operatorname{Re}}
\definecolor{mycolor1}{rgb}{0.00000,0.44700,0.94100}%
\definecolor{mycolor2}{rgb}{0.85000,0.32500,0.09800}%
\definecolor{mycolor3}{rgb}{0.92900,0.69400,0.12500}

\begin{document}

\title{Modeling of Nonlinear Interference Power for Dual-Polarization 4D Formats} 

\author{Gabriele Liga$^1$, Bin Chen$^{1,2}$, Astrid Barreiro$^1$, and Alex Alvarado$^1$}
\address{$^1$Department of Electrical Engineering, Eindhoven University of Technology, Eindhoven, The Netherlands\\
$^2$School of Computer Science and Information Engineering, Hefei University of Technology, Hefei, China}
\email{g.liga@tue.nl}

\copyrightyear{2021}

\begin{abstract}
We assess the accuracy of a recently introduced nonlinear interference model for general dual-polarization 4D formats.~ Unlike previous models for polarization-multiplexed 2D formats, an average gap from split-step Fourier simulations within 0.1 dB is demonstrated. 
\end{abstract}

\section{Introduction}
\vspace{-.15cm}
Nonlinear interference (NLI) modeling in optical fiber transmission is a key tool to analyze and optimize the performance of optical communication systems. In recent years, remarkable progress has been made in this direction, and as new model's applications emerge, further work is required to meet new accuracy and computational requirements. 
Constellation shaping is today a popular area of application for NLI models for two reasons: i) the performance of a given constellation significantly depends on the amount of NLI power induced during propagation; ii) NLI models can provide easy-to-compute as well as accurate cost functions for the performance optimization. However, as research focus moves towards multidimensional constellation shaping, where the different dimensions are mapped onto the degrees of freedom of the fiber channel (quadratures, polarization, wavelengths, etc),  extensions of the established NLI models for two-dimensional (2D) formats are needed.

NLI power models such as the enhanced Gaussian noise (EGN) model \cite{Carena2014} introduced analitical expressions explicitly linking the properties of the transmitted constellation to the resulting NLI power after propagation. Nonetheless, such expressions have been developed only for so-called polarization-multiplexed 2D (PM-2D) formats, i.e. when a single 2D constellation is used to independently map information over 2 orthogonal polarization modes of the optical field. The resulting dual polarization four-dimensional (DP-4D) constellation is, thus, given by the Cartesian product of the component 2D constellation by itself. However, the class of PM-2D formats represents only a limited subset of all possible DP-4D constellations. A vast literature on assessing the performance of DP-4D formats that do not fall within the PM-2D class is available, (see, e.g.,~\cite{Agrell09}), and, lately, DP-4D constellations have shown improved shaping gains compared to other conventional PM-2D formats \cite{Kojima2017, Chen2019, Chen2020}. However, no NLI model has been so far available to support the design of nonlinearity-tolerant DP-4D formats.  

Recently, in \cite{Liga2020}, we extended the model in \cite{Carena2014} to account for the entire DP-4D class of modulation formats. In this contribution, we present a first numerical validation of our 4D model. Moreover, we show that heuristic extensions of the EGN model to non PM-2D formats may lead to inaccuracies in the prediction of the NLI power coefficient beyond 1 dB, even for fairly regular 4D modulation formats. Our 4D model \cite{Liga2020} is instead proven to be accurate on average within 0.1 dB for all 4D modulation formats studied in this work.

\section{Analytical Formulation of the NLI coefficient for DP-4D formats}
\vspace{-.15cm}
The model we aim to validate in this work consists of an analytical formula for the computation of the vector $(\sigma^{2}_x,\sigma^{2}_y)$, where $\sigma^{2}_x$ and $\sigma^{2}_y$ represent the NLI power within the received signal bandwidth over the $x$- and $y$ polarizations, respectively. The corresponding NLI power coefficient vector $(\eta_x,\eta_y)\triangleq(\sigma^{2}_x,\sigma^{2}_y)/P^3$, where $P$ denotes the transmitted signal power. The model's expression was derived using a first-order perturbational approach under the hypothesis of single-channel transmission with quasi-rectangular pulse spectrum. A comprehensive discussion on the assumptions of the model as well as its mathematical derivation can be found in \cite{Liga2020}.   

Based on \cite[eqs.~(42)-(43)]{Liga2020}, the NLI power coefficient for the $x$ polarization can be found as
\begin{align}\label{eq:4D_NLI_formula}
\begin{split}
\eta_x&=\left(\frac{8}{9}\right)^2\frac{\gamma^2}{P^3}\left[R_s^3\left(\Phi_{1}\overline{\rchi}_1+\Phi_{2}\overline{\rchi}_2+\Phi_{3} \overline{\rchi}_3\right)+R_s^2\left(\Psi_1\overline{\rchi}_4+2\Re\{\Psi_2\overline{\rchi}_5+\Psi_3\overline{\rchi}_5^*\}+\Psi_4\overline{\rchi}_6\right.\right.\\
&\left.\left.+2\Re\{\Lambda_1\overline{\rchi}_7+\Lambda_2\overline{\rchi}_7^*(f)\}+\Lambda_3\overline{\rchi}_8+2\Re\{\Lambda_4\overline{\rchi}_9+\Lambda_5\overline{\rchi}_9^*\}+\Lambda_6\overline{\rchi}_{10}\right)+R_s\Xi_1\overline{\rchi}_{11}\right],
\end{split}
\end{align}
where $\gamma$ is the fiber nonlinearity coefficient, $R_s$ is the symbol rate and the coefficients $\Phi_1$, $\Phi_2$,  $\Phi_3$, $\Psi_1$, $\Psi_2$, \dots, $\Psi_4$, $\Lambda_1$, $\Lambda_2$, \dots , $\Lambda_6$, $\Xi_1$ are functions of several different intra- and cross- polarization moments of the DP-4D transmitted modulation format, and are given in \cite[Table 8]{Liga2020}. The $\overline{\rchi}_i$ coefficients are obtained integrating over the channel bandwidth (see \cite[eq.~(42)-(43)]{Liga2020}) the frequency-dependent integrals $\rchi_i(f)$, $i=1,2,\dots,11$ in \cite[Table 8]{Liga2020}. The expression for $\eta_y$ can be found from \eqref{eq:4D_NLI_formula} by simply applying the transformation $x\rightarrow y$ and $y\rightarrow x$. As discussed in \cite[Sec.~8]{Liga2020}, \eqref{eq:4D_NLI_formula} reduces to the EGN formula for conventional PM-2D formats. 


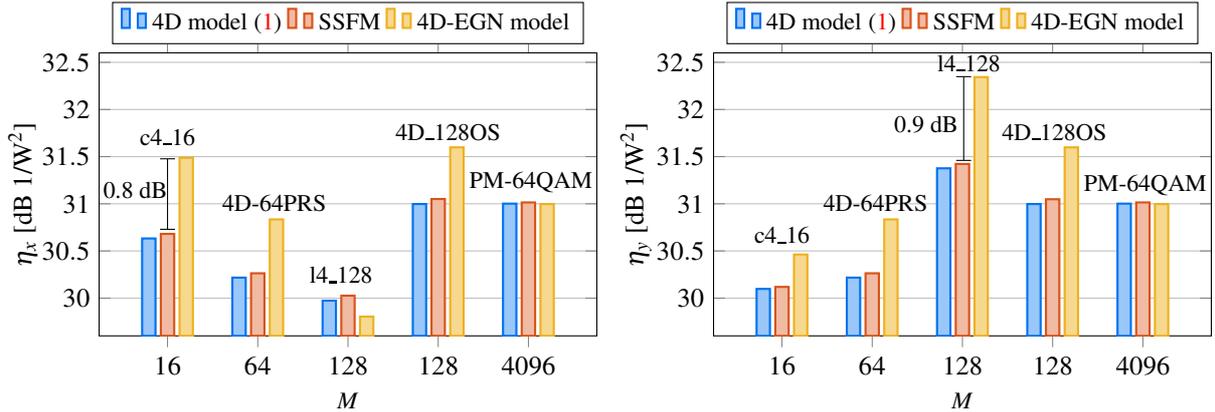
\begin{figure}[t]
\hspace{-.3cm}
\subfloat{\begin{tikzpicture}
\begin{axis}[
	ybar,
    width=3.2in,
    height=2.1in,
    enlarge x limits={abs=.9cm},
    bar width=5pt,
    ymin=29.6,ymax=32.6, 
    legend style={at={(0.5,1.18),font=\small},
    anchor=north,legend columns=-1},
    ylabel={$\eta_x$ [dB 1/W$^{2}$]},
    symbolic x coords={c416x, e, 4d64PRSx, e, l4128x, e, 4D_128OSx, e, qam4096x},
    xticklabels={,16,64,128,128,4096},
    ytick={30,30.5,31,31.5,32,32.5},
    ymajorgrids = true,
    ylabel shift={-5pt},
    xlabel={$M$},
    xlabel style={font=\small}
]
\addplot[color=mycolor1,fill=mycolor1!40,thick]
coordinates {(c416x,30.6326) (4d64PRSx, 30.2175) (l4128x,29.9732)
(4D_128OSx,30.9978)
(qam4096x,31.0018)};

\addplot[color=mycolor2,fill=mycolor2!40,thick] 
	coordinates {(c416x,30.6809) (4d64PRSx,30.2633) (l4128x,30.0262)
	(4D_128OSx,31.0508)
	(qam4096x,31.0144)};

\addplot[color=mycolor3,fill=mycolor3!50,thick] 
	coordinates {(c416x,31.4873) (4d64PRSx, 30.8348)  (l4128x,29.8042)
	(4D_128OSx,31.5987)
	(qam4096x, 30.9974)};	
	
\legend{4D model \eqref{eq:4D_NLI_formula}, SSFM, 4D-EGN model}
\node[font=\small] at (0pt,75pt) {c4\textunderscore16};
\node[font=\small] at (40pt,50pt) {4D-64PRS};
\node[font=\small] at (65pt,23pt) {l4\textunderscore128};
\node[font=\small] at (105pt,77pt) {4D\textunderscore128OS};
\node[font=\small] at (136pt,58pt) {PM-64QAM};
\draw[|-|](0pt,40pt)--(0pt,67pt);
\node[font=\small] at (-12pt,55pt) {0.8 dB};
\end{axis}
\end{tikzpicture}}\hspace{.1cm}%
\subfloat{\begin{tikzpicture}
\begin{axis}[
	ybar,
    width=3.2in,
    height=2.1in,
    enlarge x limits={abs=.9cm},
    bar width=5pt,
    ymin=29.6,ymax=32.6, 
    legend style={at={(0.5,1.18),font=\small},
    anchor=north,legend columns=-1},
    ylabel={$\eta_y$ [dB 1/W$^{2}$]},
    symbolic x coords={c416y, e, 4d64PRSy, e, l4128y, e, 4D_128OSy, e, qam4096y},
    xticklabels={,16,64,128,128,4096},
    ytick={30,30.5,31,31.5,32,32.5},
    ymajorgrids = true,
    ylabel shift={-5pt},
    xlabel={$M$},
    xlabel style={font=\small}
]

\addplot[color=mycolor1,fill=mycolor1!40,thick]
coordinates {(c416y,30.0978) (4d64PRSy, 30.2175) (l4128y,31.3763) 
(4D_128OSy, 30.9978)
(qam4096y,31.0018)};

\addplot[color=mycolor2,fill=mycolor2!40,thick] 
	coordinates {(c416y, 30.1190) (4d64PRSy,30.2633) (l4128y,31.4224) 
	(4D_128OSy,31.0486)
	(qam4096y, 31.0146)};

\addplot[color=mycolor3,fill=mycolor3!50,thick]  
	coordinates {(c416y,30.4625) (4d64PRSy, 30.8348) (l4128y,32.3436) 
	(4D_128OSy,31.5987)
	(qam4096y, 30.9974)};		
	
\legend{4D model \eqref{eq:4D_NLI_formula}, SSFM, 4D-EGN model}
\node[font=\small] at (0pt,38pt) {c4\textunderscore16};
\node[font=\small] at (35pt,50pt) {4D-64PRS};
\node[font=\small] at (70pt,103pt) {l4\textunderscore128};
\node[font=\small] at (102pt,77pt) {4D\textunderscore128OS};
\node[font=\small] at (136pt,57pt) {PM-64QAM};
\draw[|-|](68pt,66pt)--(68pt,98pt);
\node[font=\small] at (54pt,80pt) {0.9 dB};
\end{axis}
\end{tikzpicture}}
\vspace{-.3cm}\caption{NLI power coefficients $\eta_x$ (a), and $\eta_y$ (b), for DP-4D constellations with different cardinality $M$.}
    \label{fig:barplot}
\end{figure}

\section{Methodology}
\vspace{-.2cm}
\begin{wraptable}{r}{0.3\textwidth}\vspace{-.75cm}
\begin{footnotesize}
\setlength{\tabcolsep}{3pt}
\begin{tabular}{c|c}
\textbf{Parameter} & \textbf{Value}  \\
\hline \hline
\multicolumn{2}{c}{\textbf{TX parameters}}\\
\hline
    Symbol rate ($R_s$) & 32 Gbaud \\
    \hline 
    No. of channels & 1 \\
    \hline
    RRC rolloff & 0.01 \% \\
    \hline 
    TX power ($P$) & $-20$ dBm \\
    \hline
    \multicolumn{2}{c}{\textbf{Fiber parameters}}\\
    \hline
    Attenuation coeff. ($\alpha$) & 0.2 (dB/km) \\
    \hline
    Dispersion par. ($D$) & 17 ps/nm/km \\
    \hline
    Nonlinear coeff. ($\gamma$) & 1.3 (W$\cdot \text{km})^{-1}$) \\
    \hline 
    \multicolumn{2}{c}{\textbf{Link parameters}}\\
\hline
Span length & 100 km \\
\hline
No. of spans & 10 \\
\hline 
 \multicolumn{2}{c}{\textbf{SSFM parameters}}\\
 \hline
Step distribution & Adaptive step \\
\hline
$\phi_{NLmax}$ & $10^{-3}$ rad \\
\hline
Sim. bandwidth & 96 GHz \\
\hline
\end{tabular}
\end{footnotesize}
\vspace{-.2cm}\caption{System parameters.}
\label{tab:sim_par}
\end{wraptable}

The numerical validation of the model in this work is performed via the estimation of the signal-to-noise ratio in split-step Fourier method (SSFM) simulations where optical nonlinearity is kept as the sole source of noise. Indeed, in such a scenario, $\mathbf{E}\approx(P_x(\text{SNR}_x\cdot P^3)^{-1}, P_y(\text{SNR}_y\cdot P^3)^{-1})$, where $P_x$, SNR$_x$, $P_y$, and SNR$_y$ are the transmitted powers and signal-to-noise ratios over $x$ and $y$ polarization, resp.
The previous formula becomes increasingly accurate as higher-order NLI terms vanish compared to the first-order one, i.e. for small values of $P$. The simulated single-channel, multi-span optical system is described in Table \ref{tab:sim_par}. 
At the transmitter, 4D symbols are jointly modulated using a root-raised cosine (RRC) pulse shape. At the receiver, chromatic dispersion compensation and matched filtering followed by sampling are performed. $\text{SNR}_x$ and $\text{SNR}_y$ are then computed via a data-aided approach. 
 Finally, an adaptive step size SSFM with a maximum nonlinear phase rotation $\phi_{NLmax}$ per step is used to simulate the fiber propagation.

The EGN model is also used as a reference to show the increased accuracy of \eqref{eq:4D_NLI_formula}. However, in its standard formulation, the EGN model does not provide a suitable expression to account for general DP-4D formats.
An intuitive way to extend the EGN model to general DP-4D formats is to consider the EGN expressions as separately applicable to the two 2D constellations obtained by the projection of the transmitted DP-4D format over the $x$ and $y$ polarization plane. In the following, we adopt this approach, labelled 4D-EGN model, to compute EGN-based estimates of \textbf{E}. 

\section{Results}
\vspace{-.18cm}
The numerical results in this section are based on a set of DP-4D formats which combines the full list of 4D constellations in  \cite{AgrellCodesSE} (4D sphere packings), and formats used in optical communications such as the 4D 64-ary polarization ring switched (4D-64PRS) format \cite{Chen2019}, the 4D orthant-symmetric 128 (4D-OS128) format \cite{Chen2020}, and the family of 4D 2-amplitudes 8 phase-shift keying (4D-2A8PSK) formats \cite{Kojima2017}.

In Fig.~\ref{fig:barplot}, $\eta_x$ (Fig.~\ref{fig:barplot}(a)) and $\eta_y$ (Fig.~\ref{fig:barplot}(b)) are shown for some of the above mentioned constellations with different constellation cardinalities $M$, using: i) \eqref{eq:4D_NLI_formula} (blue bars); ii) the 4D-EGN model (yellow bars); iii) the SSFM (red bars). For all constellations shown, our 4D model is within 0.1 dB from the SSFM estimates for both $\eta_x$ and $\eta_y$. The 4D-EGN model leads to inaccuracies of up to 0.9 dB, for $\eta_y$ in ``l4\textunderscore128'', or 0.8 dB for $\eta_x$ in ``c4\textunderscore16''. Neither ``l4\textunderscore128'' nor ``c4\textunderscore16'' are symmetric constellations with respect to the $y=x$ plane, and this is reflected by uneven values of $\eta_x$ and $\eta_y$. However, even for symmetric constellations such as 4D-64PRS and 4D-128OS, the 4D-EGN model deviates from SSFM estimates by approx.~0.6 dB in both $\eta_x$ and $\eta_y$. As expected, for a PM-2D quadrature amplitude modulation (QAM) format such as PM-64QAM, all three estimation methods are in perfect agreement. 

To further validate \eqref{eq:4D_NLI_formula}, the average deviation $\overline{\Delta\eta}\triangleq(\Delta\eta_x+\Delta\eta_y)/2$ is computed, where $\Delta\eta_x$ and $\Delta\eta_y$ are the deviations (in dB and in absolute value) for $\eta_x$ and $\eta_y$, respectively, between a given model and the SSFM estimates.
In Fig.~\ref{fig:ensamble_accuracy}, $\overline{\Delta\eta}$ is illustrated for \eqref{eq:4D_NLI_formula} and the 4D-EGN model as a function of the constellation cardinality $M$. The solid lines show the average $\overline{\Delta\eta}$ across all considered constellations, whereas dashed lines indicate the maximum and minimum deviation. The results show an average $\overline{\Delta\eta}$ within 0.1 dB for \eqref{eq:4D_NLI_formula} across all cardinalities, with maximum $\overline{\Delta\eta}$ of 0.25 dB (for $M=16$). On the contrary, the 4D-EGN model shows deviations in excess of 1 dB for low cardinality formats ($M\leq$64), with worst-case scenario $\overline{\Delta\eta}$ of 1.75 dB for $M=16$. At higher cardinalities ($M\geq$64), the 4D-EGN model accuracy improves as the average $\overline{\Delta{\eta}}$ lies between 0.3 dB and 0.5 dB.      

\begin{wraptable}{r}{0.35\textwidth}\vspace{-.3cm}
\setlength{\tabcolsep}{3pt}
\begin{footnotesize}
\begin{tabular}{c|c|c}
\textbf{Const.~label} & $M$ & $(\eta_x,\eta_y)$ [dB 1/W$^2$] \\
\hline\hline
dicyclic4\textunderscore16 \cite{AgrellCodesSE} & 16 & (30.2, 30.2) \\
      4D-2A-8PSK5b \cite{Kojima2017} & 32 &  (30.3, 30.3) \\
      4D-2A-8PSK6b \cite{Kojima2017} & 64 &  (30.3, 30.3) \\
      4D-2A-8PSK7b \cite{Kojima2017} & 128 & (30.3, 30.3) \\
      w4\textunderscore256 \cite{AgrellCodesSE} & 256 & (30.7, 30.7)\\
      sphere4\textunderscore512 \cite{AgrellCodesSE} & 512 &(30.7, 30.7) \\
      120cell4\textunderscore600 \cite{AgrellCodesSE} & 600 & (30.3, 30.3)\\
      a4\textunderscore2048 \cite{AgrellCodesSE} & 2048 & (30.7, 30.8)\\
      a4\textunderscore4096 \cite{AgrellCodesSE} & 4096 & (30.8, 30.7)\\
      \hline\hline
\end{tabular}
\end{footnotesize}
\vspace{-.2cm}\caption{$\overline{\eta}$-optimal formats in Fig.~\ref{fig:opt}.}\label{tab:opt_const}
\end{wraptable}

Finally, in Fig.~\ref{fig:opt}, the minimum values of the average NLI power coefficient $\overline{\eta}\triangleq (\eta_x+\eta_y)/2$ are shown for each $M$ within the set of constellations analyzed in this work. The corresponding $\overline{\eta}$-optimal constellations are listed in Table \ref{tab:opt_const} for some values of $M$. The 4D-EGN model (yellow markers) almost always overestimates $\overline{\eta}$, with deviations up to 0.75 dB ($M=81$). Conversely, \eqref{eq:4D_NLI_formula} (blue markers) is consistently within 0.1 dB from the SSFM $\overline{\eta}$ (red markers). Among these optimal constellations, we highlight: i) ``w4\textunderscore256'' and ``a4\textunderscore4096", already studied in the context of coded modulation (see refs. in \cite{AgrellCodesSE}), which outperform PM-16QAM and PM-64QAM, respectively; ii) ``120cell600'' (see $x$ and $y$ projections in the inset in Fig.~\ref{fig:opt}), which is a constant-modulus constellation in 4D, and has the lowest $\overline{\eta}$ among all formats with $M\geq$128.   

\begin{figure}[t]
\begin{floatrow}
\captionsetup{singlelinecheck=false,margin={.2cm,0cm}}
\hspace{-0.3cm}\ffigbox[\FBwidth]{\definecolor{mycolor3}{rgb}{0.92900,0.69400,0.12500}%
\definecolor{mycolor1}{rgb}{0.00000,0.44700,0.94100}%
\definecolor{mycolor2}{rgb}{0.85000,0.32500,0.09800}%
\begin{tikzpicture}
\begin{axis}[%
width=2.4in,
height=1.9in,
scale only axis,
xmode=log,
xmin=16,
xmax=4096,
xtick={ 16,32,64,128,256,512,1024,2048,4096},
xticklabels={16, 32, 64, 128, 256, 512,1024, 2048, 4096},
ytick={0.1,0.5,1,1.5,1.8,2},
xminorticks=true,
xlabel style={font=\small},
xlabel={$M$},
ymin=0,
ymax=2,
ylabel shift=-5pt,
ylabel style={font=\small},
ylabel={$\overline{\Delta\eta}\text{ [dB]}$},
axis background/.style={fill=white},
xmajorgrids,
xminorgrids,
ymajorgrids,
legend pos={north east},
legend style={font=\small},
legend cell align={left},
ylabel shift={-8pt}
]
\addplot [color=mycolor3, thick, mark=triangle*, mark options={fill=white,solid}, mark size=3pt]
  table[row sep=crcr]{%
8	1.16098554574825\\
16	0.778825497202825\\
32	0.516905500205035\\
64	0.389938169860091\\
128	0.42457963690718\\
256	0.346319859210235\\
512	0.416006685275838\\
1024	0.47086157651573\\
2048	0.31216292150934\\
4096	0.241387224704972\\
};
\addlegendentry{4D-EGN model}

\addplot [color=mycolor1, thick,mark=*, mark options={fill=white,solid}, mark size=2pt]
  table[row sep=crcr]{%
8	0.0354204758182281\\
16	0.0806760105344528\\
32	0.101026065340123\\
64	0.0516168602599386\\
128	0.0419967346147656\\
256	0.0410587189928201\\
512	0.034204625126641\\
1024	0.0460699025723912\\
2048	0.0410266242512263\\
4096	0.0273501518077248\\
};
\addlegendentry{4D model \eqref{eq:4D_NLI_formula}}

\addplot [color=mycolor1, dashed, name path=A,thick]
  table[row sep=crcr]{%
8	0.0693856996427371\\
16	0.218013395957714\\
32	0.178656251312585\\
64	0.106690600996076\\
128	0.0534094943127386\\
256	0.0564194519707151\\
512	0.0374928981697629\\
1024	0.0460699025723912\\
2048	0.0428667963373091\\
4096	0.0441175294172478\\
};
\addplot [color=mycolor1, dashed, name path=B,thick,thick]
  table[row sep=crcr]{%
8	0.0224390127411667\\
16	0.0277239238897344\\
32	0.031126081840533\\
64	0.0305008210975615\\
128	0.0307047737248638\\
256	0.0296459525859838\\
512	0.0290530106470293\\
1024	0.0460699025723912\\
2048	0.0391864521651435\\
4096	0.0105827741982019\\
};

\addplot[mycolor1!40, opacity=0.3] fill between[of=A and B];

\addplot [color=mycolor3, dashed, name path=C, thick]
  table[row sep=crcr]{%
8	1.94065296338262\\
16	1.7678290042915\\
32	0.651657771349083\\
64	0.578239781270412\\
128	0.571626447220103\\
256	0.538225816229353\\
512	0.543781102618599\\
1024	0.47086157651573\\
2048	0.479467260650567\\
4096	0.471710174033859\\
};
\addplot [color=mycolor3, dashed, name path=D, thick]
  table[row sep=crcr]{%
8	0.0224408883481928\\
16	0.0441229702538166\\
32	0.309811550571037\\
64	0.200869728594416\\
128	0.0327917117140366\\
256	0.0454457571487197\\
512	0.135141302115109\\
1024	0.47086157651573\\
2048	0.144858582368112\\
4096	0.0110642753760857\\
};
\addplot[mycolor3!40,opacity=0.4] fill between[of=C and D];
\node[font=\small] (max) at (axis cs:128,1.1) {Max.~$\overline{\Delta\eta}$};
\node[font=\small] (min) at (axis cs:500,.7) {Min.~$\overline{\Delta\eta}$};
\draw[->] (max)--(axis cs:60,0.6);
\draw[->] (min)--(axis cs:700,0.28);
\end{axis}
\end{tikzpicture}%
}{\vspace{-.3cm}\caption{Average NLI power coeff. gap across the two polarizations ($\overline{\Delta\eta}$) as a function of the cardinality $M$.}\label{fig:ensamble_accuracy}}\hspace{-.6cm}
\captionsetup{singlelinecheck=false,margin={1cm,0cm}}
\ffigbox[\FBwidth]{\input{BestFormats_Eta_vs_M.tikz}}{\vspace{-.3cm}\caption{Minimum values of $\overline{\eta}$ vs. cardinality $M$ for \\ the 4D constellations investigated in this work.}\label{fig:opt}}
\end{floatrow}
\end{figure}

\section{Conclusions}
\vspace{-.2cm}
We tested a novel model predicting the NLI power for general DP-4D constellations. Based on this preliminary study, the assessed model shows a superior accuracy compared to a heuristic 4D extension of the EGN model. In particular, a 0.1 dB average deviation from SSFM simulations is demonstrated for the NLI power coefficient over a wide variety of regular 4D formats. Although further numerical validation is required, we foresee the model in this work as a powerful analytical tool for optimizing constellations in the DP-4D space of the optical field.




\vspace{.15cm}
 \begin{footnotesize}
\textbf{Acknowledgements:} {\setstretch{0.5} The work of G.~Liga is funded by the EuroTechPostdoc programme under the European Union’s Horizon 2020 research and innovation programme (Marie Skłodowska-Curie grant agreement No. 754462). This work has received funding from the European Research Council (ERC) under the European Union's Horizon 2020 research and innovation programme (grant agreement No. 757791).\par}
\end{footnotesize}
\vspace{-.3cm}
\bibliographystyle{osajnl.bst}
\bibliography{Biblio.bib}

\begin{thebibliography}{1}
\newcommand{\enquote}[1]{``#1''}

\bibitem{Carena2014}
A.~Carena, G.~Bosco, V.~Curri, Y.~Jiang, P.~Poggiolini, and F.~Forghieri,
  \enquote{{EGN} model of non-linear fiber propagation,}
  {\protect\JournalTitle{Opt. Express}} \textbf{22}, 16335--16362 (2014).

\bibitem{Agrell09}
E.~Agrell and M.~Karlsson, \enquote{Power-efficient modulation formats in
  coherent transmission systems,} {\protect\JournalTitle{JLT}} \textbf{27},
  5115--5126 (2009).

\bibitem{Kojima2017}
K.~{Kojima}, T.~{Yoshida}, T.~{Koike-Akino}, D.~S. {Millar}, K.~{Parsons},
  M.~{Pajovic}, and V.~{Arlunno}, \enquote{Nonlinearity-tolerant
  four-dimensional {2A8PSK} family for 5–7 bits/symbol spectral efficiency,}
  {\protect\JournalTitle{JLT}} \textbf{35}, 1383--1391 (2017).

\bibitem{Chen2019}
B.~{Chen}, C.~{Okonkwo}, H.~{Hafermann}, and A.~{Alvarado},
  \enquote{Polarization-ring-switching for nonlinearity-tolerant geometrically
  shaped four-dimensional formats maximizing generalized mutual information,}
  {\protect\JournalTitle{JLT}} \textbf{37}, 3579--3591 (2019).

\bibitem{Chen2020}
B.~Chen, A.~Alvarado, S.~van~der Heide, M.~v.~d. Hout, H.~Hafermann, and
  C.~Okonkwo, \enquote{Analysis and experimental demonstration of
  orthant-symmetric four-dimensional 7 {bit/4D-sym} modulation for optical
  fiber communication,} {\protect\JournalTitle{arXiv:2003.12712}}  (2020).

\bibitem{Liga2020}
G.~Liga, A.~Barreiro, H.~Rabbani, and A.~Alvarado, \enquote{Extending fibre
  nonlinear interference power modelling to account for general
  dual-polarisation {4D} modulation formats,} {\protect\JournalTitle{Entropy}}
  \textbf{22}, 1324 (2020).

\bibitem{AgrellCodesSE}
\enquote{Sphere packings of dimension 4,}
  \url{https://codes.se/packings/4.htm}.

\end{thebibliography}







\end{document}